\documentclass[12pt,letterpaper]{article}
\usepackage[margin=1in]{geometry}
\usepackage{color}
\usepackage{comment}
\usepackage[numbers]{natbib}
\usepackage{graphicx}
\usepackage{amsmath}
\usepackage{amssymb}
\usepackage{subfigure}
\usepackage{lineno}
\usepackage{hyperref}
\setcitestyle{square}

\usepackage{sectsty}
\sectionfont{\large}
\subsectionfont{\normalsize}
\subsubsectionfont{\normalsize}
\paragraphfont{\normalsize}

\begin{document}

\pagenumbering{gobble}
\huge
\begin{center}
%\title{%
Astro2020 Science White Paper\\
\vspace{0.1in}

Deep Multi-object Spectroscopy to Enhance Dark Energy Science from LSST
\end{center}
\normalsize

%\author{}%The LSST Dark Energy Science Collaboration}
%\maketitle
\noindent\textbf{Thematic Areas:} Cosmology and Fundamental Physics\\

\noindent\textbf{Principal (corresponding) author:}\\
Name: Jeffrey A. Newman\\
Institution: University of Pittsburgh and PITT PACC\\
Email: janewman@pitt.edu \\
Phone: (412) 592-3853\\

\noindent\textbf{Co-authors:} J.~Blazek (SNSF Ambizione, EPFL; CCAPP, Ohio State U.), N.~E.~Chisari (Oxford), D.~Clowe (Ohio U.), I.~Dell'Antonio (Brown), E.~Gawiser (Rutgers), R.~A.~Hlo\v{z}ek (Toronto), A.~G.~Kim (LBNL), A.~von der Linden (Stony Brook University), M.~Lochner (African Institute for Mathematical Sciences, South African Radio Astronomy Observatory), R.~Mandelbaum (CMU), E.~Medezinski (Princeton), P.~Melchior (Princeton), F.~J.~S\'{a}nchez (UC Irvine), S.~J.~Schmidt (UC Davis), S.~Singh (UC Berkeley, BCCP), and R.~Zhou (U. Pittsburgh, PITT PACC),for the LSST Dark Energy Science Collaboration\\

%\begin{abstract}

\noindent\textbf{Abstract:} Community access to deep ($i \sim 25$), highly-multiplexed optical and near-infrared multi-object spectroscopy (MOS) on 8--40m telescopes would greatly improve measurements of cosmological parameters from LSST. 
%Every cosmological probe that we plan to apply to LSST data would benefit from these capabilities. 
%Every cosmological probe that we plan to apply to LSST data within the LSST Dark Energy Science Collaboration would benefit from these capabilities.  
The largest gain would come from improvements to LSST photometric redshifts, which are employed directly or indirectly for every major LSST cosmological probe; deep spectroscopic datasets will enable reduced uncertainties in the redshifts of individual objects via optimized training.   Such spectroscopy will also determine the relationship of galaxy SEDs to their environments, key observables for studies of galaxy evolution.  The resulting data will also constrain the impact of blending on photo-$z$'s. Focused spectroscopic campaigns can also improve weak lensing cosmology by constraining the intrinsic alignments between the orientations of galaxies. %; this effect can be investigated in conjunction with photometric redshift training.  
Galaxy cluster studies can be enhanced by measuring motions of galaxies in and around clusters and by testing photo-$z$ performance in regions of high density. Photometric redshift and intrinsic alignment studies are best-suited to instruments on large-aperture telescopes with wider fields of view (e.g., Subaru/PFS, MSE, or GMT/MANIFEST) but cluster investigations can be pursued with smaller-field instruments (e.g., Gemini/GMOS, Keck/DEIMOS, or TMT/WFOS), so deep MOS work can be distributed amongst a variety of telescopes. However, community access to large amounts of nights for surveys will still be needed to accomplish this work.  In two companion white papers we present gains from shallower, wide-area MOS and from single-target imaging and spectroscopy.%\end{abstract}

\newpage
\pagenumbering{arabic}

\section{Introduction}

%
%\textbf{Currently unassigned.  Should introduce the basic idea that this is about supporting/enabling dark energy science with LSST, with a variety of science cases listed below that would benefit from multi-object spectroscopy.}
%
%\done{(entire section)}

The Large Synoptic Survey Telescope (LSST) will play a major role in improving our knowledge of cosmology over the years 2023--2033, constraining fundamental cosmological parameters using multiple complementary methods.  %However, obtaining measurements that extract the full potential of LSST will require additional data from other ground-based facilities.  
However, the baseline dark energy analyses that will be carried out by the LSST Dark Energy Science Collaboration (DESC) will require additional data from other ground-based facilities to improve photometric redshift estimates, reduce systematic uncertainties, and realize the full potential of LSST \cite{descsrd}.  

In this white paper, we describe the science opportunities to enhance cosmological measurements from LSST that would be enabled by community access to {\bf deep ($i \sim 25$), highly-multiplexed} optical and near-infrared multi-object spectroscopy on 8--40m telescopes. 
Every cosmological probe that we plan to apply to LSST data would benefit from these capabilities. 
%Every cosmological probe that we plan to apply to LSST data within the LSST Dark Energy Science Collaboration would benefit from these capabilities.  
%This begins with improvements to LSST photometric redshifts, which depend on deep spectroscopic datasets for optimization of algorithms; every major cosmological probe will rely on photo-$z$'s directly or indirectly.  However, focused spectroscopic campaigns can improve weak lensing cosmology by constraining the intrinsic alignments between the orientations of galaxies; galaxy cluster studies by measuring kinematics of galaxies in clusters and testing photo-$z$ performance in regions of high density; and strong lensing measurements by determining the distribution of foreground mass along the line of sight to systems.  
%We describe these opportunities in more detail below.

In two companion white papers, we describe the gains for LSST cosmology that would come from community access to wider-field multi-object spectroscopy (spanning areas $>20$ deg$^2$) and from follow-up single-target imaging and spectroscopy of supernovae and strong lens systems \cite{wide,single}.  

%
%\section{Science cases}
%
%\textbf{We should consider whether to list the science cases as below, or to group them strategically (e.g., photo-z, static science cases, transient science cases, \dots) but for now it's just a list with separate volunteers for each one.}
%
%Here's a list of tables/figures that can be integrated into the relevant subsections:
%
%\begin{itemize}
%\item Photo-z errors vs. training sample size (Zhou, Newman)
%\item Photo-z training time for different instrumentation (Newman, done)
%\item FoM for different photo-z error scenarios (Eifler, Krause)
%\item Photo-z calibration errors for different-size southern surveys (Newman)
%\item SN cosmology for different-size spectroscopic sets (Hlozek, done) 
%\item Impact on WL+LSS cosmology constraints of being able to decrease the size of priors on intrinsic alignment models due to increased spectroscopic sample size (could use DESC SRD infrastructure)
%\item Table: this is the science case, these are the facilities that could contribute
%\end{itemize}

\section{Deep Spectroscopic Samples for Photo-{\em z} Training}

Photometric redshifts are critical for all LSST probes of cosmology.  The cosmological tests we will perform all rely on determining the behavior of some quantity with redshift, $z$, but we cannot measure spectroscopic redshifts (spec-$z$'s) for the large numbers of objects detected in the LSST imaging data.    Even in cases where follow-up spectroscopy of individual objects will be needed (e.g., strong lens systems and some supernova studies), photo-$z$'s are used to identify targets of interest.  However, if photometric redshift estimates are systematically biased, dark energy inference can be catastrophically biased as well (see, e.g., \cite{hearin_etal10}); as a result photo-$z$'s are both a critical tool and a major source of concern affecting all cosmological analyses.  The great depth of LSST data is a compounding factor, as it exacerbates any challenges associated with follow-up spectroscopy to the depths of the samples used for LSST.

Lacking a comprehensive knowledge of galaxy evolution, the only way in which photo-$z$ errors can be reduced and biases characterized is via galaxies with robust spectroscopic redshift measurements.  We follow \cite{specneeds} in dividing the uses of spec-$z$'s into two  classes, ``training'' and ``calibration.'' 
%in general, calibration will be the harder problem. 
 {\bf Training} is the use of samples with known $z$ to develop or refine algorithms, and hence to {\it reduce  random errors on individual objects' photo-$z$'s}.  Photometric redshifts that are trained from larger and more complete spectroscopic samples greatly improve the constraining power of LSST, for example by providing sharper maps of the large-scale structure 
 %and stronger cosmological constraints; e.g., based upon simulations ideal training samples can reduce LSST photo-$z$ errors by $2.5\times$ and 
 and improved clustering statistics, providing better photometric classifications for supernovae, enabling identification of lower-mass galaxy clusters at higher confidence, and yielding better intrinsic alignment mitigation for weak lensing measurements.  If photo-$z$'s are limited only by photometric errors (as with a perfect training set), LSST can deliver photo-$z$ estimates with sub-$2\%$ uncertainties ($\sigma_z < 0.02(1+z)$), but errors in real data sets at LSST depth with our current knowledge of galaxy spectral energy distributions are closer to 5\%.  Achieving the ideal performance by having a large training sample spanning the properties of objects used for cosmology would improve the  Dark Energy Task Force Figure of Merit from LSST lensing and Baryon Acoustic Oscillations alone by $\sim 40\%$ \cite{detf,2006JCAP...08..008Z}.
 %The LSST survey is capable of delivering photo-$z$'s with an RMS error of $\sigma_z = 0.02(1+z)$ in simulations with perfect knowledge of the range of galaxy spectra, but actual samples of similar S/N yield $\sigma_z \sim 0.05(1+z)$.  
% With an ideal training set of redshifts, we could achieve the LSST-limited performance, 
%improve the DETF figure of merit from the combination of LSST lensing and Baryon Acoustic Oscillation measurements by $\sim 40\%$. 

{\bf Calibration} is the problem of determining the true overall $z$ distribution of a sample of objects; miscalibration will lead to {\it systematic} errors in photo-$z$'s and hence downstream analyses \cite{2006ApJ...644..663Z,2006JCAP...08..008Z,2006astro.ph..5536K,tysonconf,hearin_etal10}.  
%In general, for most dark energy science the aggregate properties (e.g., mean and standard deviation of the redshift distribution) of samples selected according to photo-$z$ information must be determined with very high accuracy for cosmological inference not to be systematically biased at uncacceptable levels; hence, high-precision, robust calibration is necessary
%Photo-$z$ calibration requirements for LSST are extremely stringent \cite{descsrd}.  
%drop17
%For LSST, for instance, it is estimated that the mean redshift for each sample used for cosmology (typically, objects selected within some bin in photometric redshift) must be known to $\sim 2\times 10^{-3} (1+z)$, i.e., 0.2\% \cite{2006ApJ...644..663Z,2006JCAP...08..008Z,2006astro.ph..5536K,tysonconf,hearin_etal10}.%; the true width of the redshift distribution of the sample must also be known to similar precision.
%
% ($\Delta \sigma_z < \sim 3\times 10^{-3} (1+z)$, where $\sigma_z$ is the Gaussian sigma of the true redshift distribution.
%If photo-z's are systematically biased in an unknown way or if their errors are poorly understood, dark energy inference will be biased as well (e.g., because we are effectively measuring the distance to a different redshift than is assumed in calculations).  
%Essentially, the goal of calibration is to determine with high accuracy the moments of the distribution of true redshifts of dark energy samples.  
For both training and calibration, we need sets of galaxies for which the true $z$ is securely known.  If spec-$z$'s could be obtained for a large, unbiased sample of objects, both needs can be fulfilled using the same data. However, many faint galaxies fail to yield secure $z$'s; hence other methods may be needed for calibration, as described in a companion paper \cite{wide}. Training still will benefit from incomplete samples.  

\begin{figure}[t]
\centering
\includegraphics[width=0.85\textwidth]{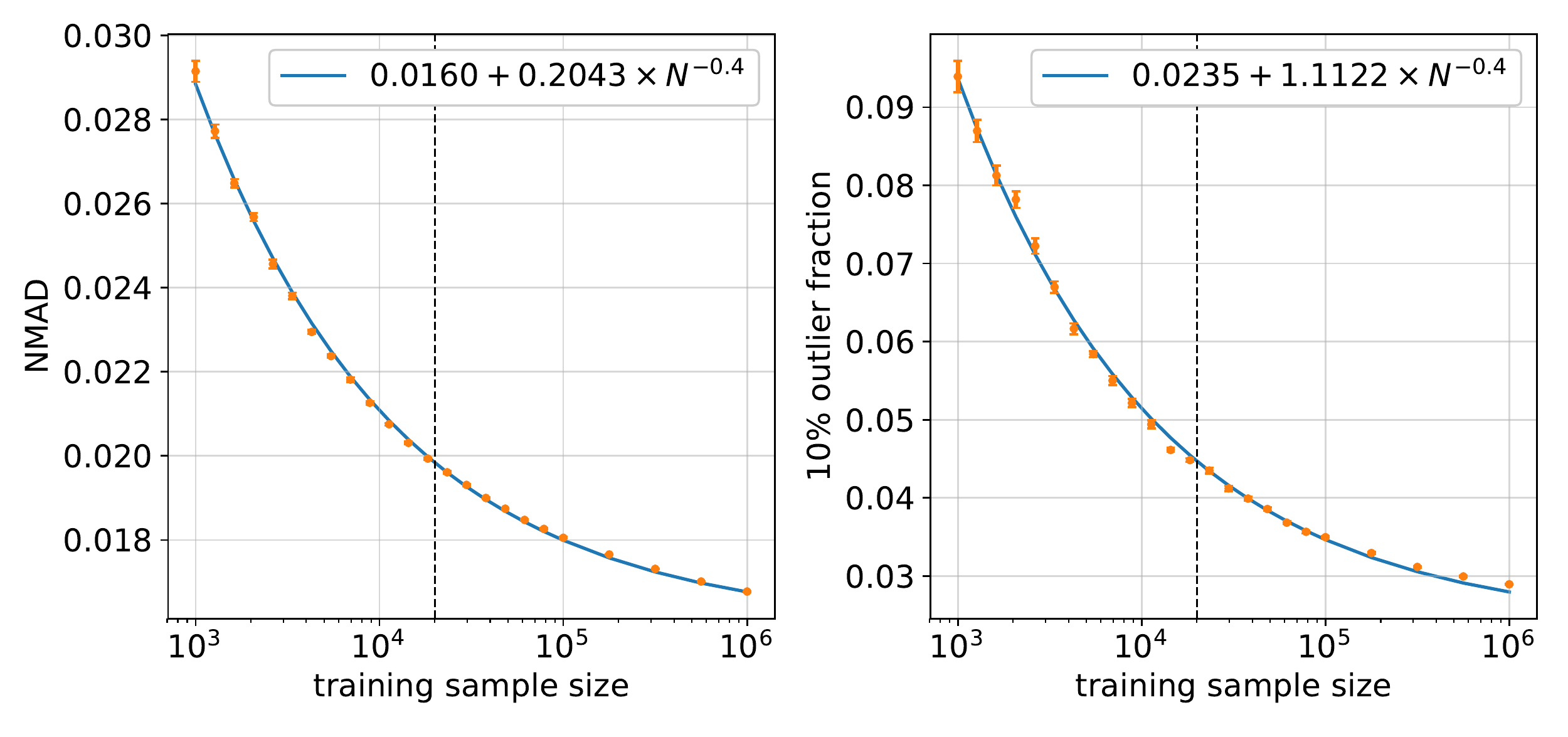}
\caption{Orange points show photometric redshift errors and outlier rates versus the number of galaxies in the training set for galaxies with simulated LSST photometric errors.   Photo-$z$'s were calculated using a random forest regression algorithm. The left panel shows the photo-$z$ error, quantified by the normalized median absolute deviation (NMAD) in $(z_\text{phot}-z_\text{spec})/(1+z_\text{spec})$, as a function of training set size; similarly, the right panel shows the fraction of 10\% outliers, i.e. objects with $|z_\text{phot}-z_\text{spec}|/(1+z_\text{spec})>0.1$. A vertical dashed line shows the sample size for the baseline training survey from \cite{specneeds}.  The blue curves represent simple fits to the measurements as a function of the training set size, $N$. This analysis uses a set of simulated galaxies from Ref.~\cite{Graham2018} that spans the redshift range of $0<z<4$, using a randomly-selected testing set of $10^5$ galaxies for estimating errors and outlier rates; these catalogs are based upon simulations from Refs. \cite{grahamcat1},\cite{grahamcat2}, and \cite{grahamcat3}.}
\label{fig:training_size}
\end{figure}

In a recent paper \cite{specneeds}, it was concluded that an effective training set of photometric redshifts for the LSST weak lensing sample would require highly-multiplexed medium-resolution ($R \sim 4000$) spectroscopy covering as much of the optical/infrared window as possible with very long exposure times on large telescopes. To enable photo-$z$ direct calibration errors to be subdominant to other uncertainties if the training set were used for that purpose, the spec-$z$ sample must comprise at least 20,000 galaxies spanning the full color and magnitude range used for cosmological studies, reaching $i=25.3$; as can be seen in Fig.~\ref{fig:training_size}, improvements in photo-$z$ errors and outlier rates are slow beyond this point.

A critical issue for machine learning-based photometric redshift  algorithms %\citet{2004PASP..116..345C,
\cite[e.g.,][]{2015MNRAS.452.3100C} is that sample/cosmic variance in the regions with spectroscopy can imprint on the redshift distribution over the whole sky, biasing photo-$z$ results.  
To both quantify and mitigate this effect, the survey strategy described in Ref.~\cite{specneeds} seeks to obtain spectroscopy spanning at least 15 widely-separated fields a minimum of 20$^\prime$ in diameter.  
Such a survey has comparable sample/cosmic variance to the Euclid C3R2  strategy of six 1 sq.~deg.~fields \cite{2017ApJ...841..111M}, but requires only $\sim$ 22\% as much sky area to be covered; by spanning more fields than C3R2 it also allows more robust identification of regions that are overdense or underdense at a given $z$.  
We have calculated the amount of dark time required for such a survey (assuming one-third losses for weather and overheads) for a variety of instruments and telescopes of varying characteristics, updated from the tables in Ref. \cite{specneeds}; the results are summarized online at \url{http://d-scholarship.pitt.edu/id/eprint/36036}.  
%Additional spectroscopy of higher-redshift clusters may be needed as these objects are rare, as described in Appendix \ref{appendix:cluster_specz}. 

%\textcolor{red}{add the figures and tables to support this.  Rongpu Zhou has made a figure of photo-z error vs. sample size.  Tim Eifler and Elisabeth Krause are working on a measure of DE constraint vs. photo-z error.  There should be two tables, one of instrument characteristics and one of survey times.  Maybe the latter could be an appendix/addendum past the 5 page limit, if we need?}

We note that such a survey has strong synergies with studies of galaxy evolution: it would determine the range of galaxy SEDs (and hence star formation histories) as a function of their local environment across the redshift and magnitude range covered by LSST cosmology samples (reaching down to $\sim L_*$ at $z=2$, and including brighter galaxies to $z=3$).  Improved photo-$z$'s will also enhance a variety of galaxy evolution science from LSST.

\subsection{Testing the Impact of Blending on Photometric Redshifts}

%\textbf{Volunteers: Melchior, Sanchez}

%Compared to current surveys like DES with $r \approx 24.3$~\citep{DES16}, the increasing coadd depth of LSST at $r \sim 27.5$~\citep{Ivezic08} will yield a higher number density of detected sources. In addition, every source will have a larger area over which its light emission is brighter than the sky background. Consequently, t
Due to its unprecedented depth and sensitivity to the low-surface-brightness outer regions of galaxies, the probability of two or more sources overlapping in LSST data is high. 
Approximately 63\% of LSST sources will have at least 2\% of their flux coming from other objects, in contrast to $\sim$30\% of  DES sources~\citep{Samuroff18,Sanchez19}. 
These overlaps complicate the measurement of galaxy fluxes and shapes, requiring accurate deblending \citep{Melchior18,Euclid19} or statistical correction techniques. Any residual light contamination will result in biases of individual-object photo-$z$'s or potentially even in estimates of overall redshift distributions~\citep{Gruen18}.

Deep multi-object, medium-resolution spectroscopy can detect the presence of blends and the redshifts of each component by identifying superimposed features in a spectrum, providing new training samples for deblenders and constraints on statistical corrections for deblending effects.  While many blends can be detected as having multiple components in space-based data, some are so close that spectroscopy will provide the only definitive indication of a problem.  As a result, the proposed photometric redshift training spectroscopy should also greatly enhance studies of deblending down to the depth of the LSST Gold sample, $i_\text{lim}=25.3$~\citep{descsrd}.
%Source deblending is therefore an important field of ongoing work, e.g. through improved source separation or machine learning techniques~\citep{Melchior18,Euclid19}. To reliably train machine learning methods, the sample with known high-quality spectra needs to mimic the sample used for cosmological studies. Thus, deep multi-object, high-resolution spectroscopy can significantly increase the currently limited training sample to the depth of the LSST Gold sample, $i_\text{lim}=25.3$~\citep{descsrd}.

%Also for a statistical correction of blending effects, it is key to understand whether the blended constituents are at different redshifts. In existing multi-object spectroscopic campaigns, a few percent of the targets are blended and yield redshift solution for multiple constituents of a blend. Deep multi-object spectroscopy will likely have a higher rate of blended objects and could ideally provide multiple redshift solutions to the depth of the LSST Gold sample. This will be particularly important but challenging for the strongly clustered early-type galaxies that lack strong emission lines.

%\todo{Tim Eifler is still working on calculations of constraints vs. photo-z errors.}

\section{Constraining Models of Intrinsic Galaxy Alignments}
%\done{}
%\textbf{Volunteers: Chisari, Blazek, Singh. NOTE: This could also be part of the photo-z training section.}  

As discussed in the companion white paper on wide MOS, intrinsic correlations between galaxy shapes (``intrinsic alignments'' or IA) induced by local/environmental effects are an important contaminant to weak lensing measurements; some constraints on this effect can be obtained via cross-correlation measurements, as described there.
%While wide MOS observations will provide valuable IA measurements, they will not extend to the fainter population that will dominate the LSST lensing sample.
By enabling the 3D localization of galaxies, a deep MOS campaign would provide greatly-improved direct constraints on intrinsic alignments for {typical} weak lensing sources, rather than only for the bright and nearby objects which current datasets constrain \citep{Joachimi11,Samuroff18,Johnston18}.  
%significantly increasing the current range in galaxy redshift and luminosity compared to current constraints and what is accessible from future wide MOS. The redshift evolution and luminosity dependence of the alignment signal remain poorly constrained in the regime relevant for weak lensing studies. 
%Furthermore, it is unclear whether blue/disk-like galaxies exhibit significant IA -- cosmological hydrodynamical simulations currently yield conflicting results . 
Such data would extend our knowledge of IA to unexplored regimes, resolve the current inconsistencies between predictions of different hydrodynamical simulations \citep{Tenneti15,Chisari15,Chisari16}, and  allow better  priors to be placed on IA parameters, increasing the cosmological constraining power of LSST.

%Finally, galaxy shapes and angular momenta are closely related to the underlying processes of galaxy and halo formation, and deep MOS observations will improve our understanding of the acquisition of galactic angular momentum and shapes. [How deep can we realistically get if we need ~100k galaxies at z $\sim$ 1? How small an area can we use if we want to probe NLA vs small scale behavior?]

Currently, the faintest magnitude-limited sample with an IA detection has $i_\text{lim}<19.8$ \cite{Johnston18}.  Forecasts based on recent measurements \cite{Singh14,Johnston18} indicate that meaningful IA constraints require $\gtrsim 10^5$ galaxies with measured shapes and spectra. Obtaining this many spectra for a representative sample at the magnitude limit of the LSST Gold sample ($i_\text{lim}=25.3$) would require a significant expansion of the photo-$z$ training program described above, and may be infeasible with currently planned facilities.
%A deep field with a contiguous area of [??] deg$^2$ would allow a measurement of IA for the faint objects that will dominate the lensing source catalogs for LSST and other future lensing surveys. Reaching the magnitude limit of the LSST Gold sample ($i_\text{lim}=25.3$) is challenging, although it would be feasible with an expansion of the photo-$z$ strategy described above [...].
However, even if we cannot reach magnitudes as faint as $i_{\rm lim}=25.3$, extending direct IA tests closer to the LSST limits would be very valuable. 
Most of the IA signal will come from scales where shape noise dominates, and thus total IA constraining power (for a fixed number of galaxies) is maximized if the surface density of targets is high. Such a dense sample can be obtained by switching out targets to other, brighter galaxies during the photo-$z$ training survey once secure redshifts are obtained; in this way, $\gtrsim 10^5$ spec-$z$'s for  objects with $i \lesssim 24$ should be obtainable during the training survey (only 9\% as much observing time is needed to obtain the same S/N at $i=24$ as at $i=25.3$). Additional information will come from cross-correlations with shallower, wider-area surveys, as described in a companion paper \cite{wide}. %COMMENT NEEDED: DO YOU NEED/WANT SPACE-BASED DATA FOR THIS?   % NO
%As an example of a feasible program, we find that obtaining $\sim$200k redshifts for a sample with $i_\text{lim}<24$ {\bf [change to 23.5? Jeff, can you confirm these numbers?]} is possible with $\sim$50 PFS or 25 MSE pointings, requiring a maximum of $\sim$0.72 or 0.36 times as much time as the fiducial photo-z training survey. Some of this time can likely be obtained without additional resources by using one or more photo-z fields, especially since the relatively brighter objects can be quickly filled in while observing the fainter end of the photo-z sample.  In this example, we consider a roughly 4 sq.\ deg.\ field, which would allow us to place a conservative upper limit on the fiducial IA amplitude of $A_{IA}=1$ at $\sim3\sigma$. Nonlinear IA effects which boost the signal at small scales, will potentially be more tightly constrained.% (or potentially detected).

%Obtaining $\sim$100k redshifts for a sample with $i_\text{lim}<24$ is possible with $\sim$50 PFS or 25 MSE pointings, requiring $\sim$0.36 or 0.18 times as much time as the fiducial photo-z training survey, and covering 65 or 32 sq. deg., respectively.

\section{Enhancing Cluster Cosmology via Spectroscopy}
%Tests of photometric redshifts in cluster fields}

%\textbf{Volunteers: Medezinski, Clowe, von der Linden}

%\done{}

Deep MOS of galaxies in a set of fields containing galaxy clusters will improve LSST cluster cosmology in a number of ways, while simultaneously resolving open questions about galaxy evolution by determining differences between SEDs of galaxies in clusters versus the field.

{\bf Training and Testing Photometric Redshifts in Cluster Fields:} Photometric redshifts are critical for weak-lensing mass calibration of galaxy clusters. They are already a leading source of systematic uncertainty in current work \citep{WtG3,McClintock19}, with even more stringent requirements on photo-$z$ accuracy for LSST \citep{descsrd}. Photo-$z$'s are vital for distinguishing the lensed background galaxy population from the unlensed foreground and cluster population.
%;  unaccounted-for contamination of the background will lead to underestimation of cluster masses.  
However, photo-$z$ performance may degrade in higher-density regions due to the differing galaxy populations of clusters vs. the field, magnification and reddening of background sources, and severe blending due to cluster galaxies \citep[e.g.][]{CLASH_photoz}.  

Photo-$z$ algorithms are generally trained and evaluated on fields selected not to contain massive structures.  To ensure the robustness of cluster work, it will therefore be important to obtain additional MOS
%There is thus no a priori reason to expect photo-z algorithms to perform as well in clusters as in the field. In cluster fields, the mix of early- vs. late-type galaxies is different than in the field; in addition, the incidence of blending increases due to the denser environment and contamination by intra-cluster light, causing larger errors in the photometry .  
%Lensing magnification changes the observed brightness and size, leading to potential mismatches with the priors.  
%Astrophysically more interesting are the possibilities that, due to the profound impact of the cluster environment on galaxy evolution, clusters might host galaxy types very rare in the field, as well as the potential presence of dust in clusters, which would cause reddening of background galaxies.  For photo-z applications in clusters, it is therefore vital to quantify the performance of photo-z algorithms in cluster fields specifically.  Doing so requires comprehensive MOS 
spectroscopy to weak lensing depths for a sample of $\sim 20$ clusters spanning a range of redshifts. This is best achieved with high-throughput, high-multiplex spectrographs with FOVs of $\sim 10^\prime$ (wide-format IFUs may be suitable in cluster cores).  %This spectroscopy should improve photo-$z$ training for cluster galaxies 
%(beyond $z = 1$, photo-$z$ performance for HyperSuprimeCam clusters degrades from $\sim 1\%$ to $\sim 5\%$ due to the scarcity of spec-$z$'s for such objects) 
% JAN: commenting that out; lack of a 4000-angstrom break could also be an issue.
%while also providing tests of photo-$z$ performance and systematics in cluster regions.

Such a program would be able to characterize the performance of photo-$z$ probability distributions as a function of magnitude, redshift, and cluster properties.  A sample of 1000 objects per cluster allows a crude binning into 3 magnitude and 3 redshift bins with $
\sim$100 galaxies per bin, enough for a statistically meaningful evaluation of photo-$z$ performance at the percent level.  This should be done for a range of cluster redshifts, masses, and dynamical states, requiring ideally $>20$ clusters.  It is critical to achieve near-complete redshift success rates to avoid biases from target populations for which no redshift is measured in a first attempt.  Such data will also yield very valuable insight into deblending performance in general.

{\bf Measurements of Cluster Kinematics and Infall Velocities:}
%\textbf{Volunteers: Dell'Antonio, von der Linden}
%\done{}
Deep MOS observations in the fields of clusters will also enable direct mass estimates for galaxy clusters via the infall method \citep{Diaferio99,Rines03,Falco13,Arthur17},  providing an additional calibration of the mass-richness relation \citep{Rines18} ``for free" from the photometric redshift training/test spectroscopy.  We hope to obtain $\sim 200$ redshifts at projected separation $< 5$ Mpc for each cluster targeted for photo-$z$ studies, providing purely kinematic mass estimates.  In conjunction with weak-lensing measurements, spectroscopy of halo infall regions also enables sensitive comparison between the dynamical and the lensing potential of gravity, which can differ from each other at measurable levels in several modified gravity theories that seek to provide an alternate explanation for cosmic  acceleration \citep[e.g.,][]{Zu14}.  

%
%\section{Strong lensing cosmology}
%
%%\textbf{Volunteers: Tom Collett}
%LSST will discover 100,000 strong gravitational lenses \citep{Collett15}, 100 times more than are currently known. This step change opens new opportunities to precisely constrain dark energy with lensed quasars \citep{Bonvin17}, lensed supernovae \citep{Goldstein18}, lenses with multiple background sources \citep{Collett14}, lenses with a spectroscopic velocity dispersion \citep{Grillo08} and as a calibration tool for weak lensing shear \citep{Birrer18}. Each of these science cases requires redshifts for lens and source, both to confirm candidates as lenses and to convert lensing derived quantities into cosmologically meaningful constraints. The LSST lenses will be uniformly distributed across the LSST footprint, with typical r band magnitudes of $20.1 \pm 2.4$ for the lens and $23.7 \pm 0.7$ for the source. Piggbacking on a wide MOS survey across the LSST footprint would deliver the strong lens spectroscopic followup can only be practically achieved by piggybacking on a wide MOS survey reaching.

\section{Recommendations}

%\textbf{Currently unassigned.   At minimum, want to explain the types of facilities that we will need access to, ideally with a rough estimate of how much time is needed (i.e. few nights vs. hundreds of nights vs 10 years...). For wide: surface density, depths, areas.  For deep: survey times for different scenarios.}

Given the large gains to LSST cosmological studies that will come from deep multi-object spectroscopy, we recommend that access to modestly-wide-field, highly-multiplexed, large aperture spectroscopic facilities be pursued during the next decade.  Specifically,

\begin{itemize}

\item For photometric redshift training and tests of blending and intrinsic alignment effects, it is desirable to have an instrument of maximal multiplexing with a field of view of at least 20$^\prime$ diameter. Subaru/PFS, the Maunakea Spectroscopic Explorer (MSE), or GMT/GMACS with the MANIFEST fiber feed are all well-suited for this work (wider-field fiber-fed spectrographs on other $>$6~m telescopes could also be suitable).\footnote{See \url{http://d-scholarship.pitt.edu/id/eprint/36036} for survey time estimates for various instrument/telescope combinations.}  Personnel costs are likely to be high on smaller telescopes because of the longer time required for deep surveys on them, making instruments on 4~m telescopes such as DESI unattractive for this work.

\item For galaxy cluster studies, targeting more objects over smaller fields of view is desirable.  For that work, suitable options may include Gemini/GMACS, Keck/DEIMOS or LRIS, GMT/GMACS (in multislit mode), TMT/WFOS, or a new, higher-multiplex spectrograph at one of these observatories; wider-field facilities such as DESI, Subaru/PFS or MSE would be less well-suited.

\end{itemize}

The telescope time required for photo-$z$ training is substantial ($\sim$6 months of dark nights if one of the most optimal facilities is used exclusively), but it can be spread out over the ten-year span of the LSST survey, reducing the impact in any one year.  Additionally, these studies are highly synergistic with studies of galaxy evolution, potentially allowing combined surveys with a large impact on multiple fields of research.

\newpage
\begin{center}
{\large\bf Acknowledgements}    
    
\end{center}

The LSST Dark Energy Science Collaboration acknowledges ongoing support from the Institut National de Physique Nucl\'eaire et de Physique des Particules in France; the Science \& Technology Facilities Council in the United Kingdom; and the Department of Energy, the National Science Foundation, and the LSST Corporation in the United States.  DESC uses resources of the IN2P3 Computing Center (CC-IN2P3--Lyon/Villeurbanne - France) funded by the Centre National de la Recherche Scientifique; the National Energy Research Scientific Computing Center, a DOE Office of Science User Facility supported by the Office of Science of the U.S.\ Department of Energy under Contract No.\ DE-AC02-05CH11231; STFC DiRAC HPC Facilities, funded by UK BIS National E-infrastructure capital grants; and the UK particle physics grid, supported by the GridPP Collaboration.  This work was performed in part under DOE Contract DE-AC02-76SF00515.

\newpage
\bibliographystyle{unsrt_truncate}
\bibliography{main}

\end{document}